\documentclass[journal,twocolumn]{IEEEtran}
\usepackage{cite}
\usepackage{xcolor}
\usepackage[utf8]{inputenc} % Input encoding
\usepackage{graphicx} % For including images
\usepackage{subcaption}
\usepackage{graphicx,hyperref,subcaption}
\usepackage{amsmath,amsfonts,amssymb} % For mathematical symbols and equations
\usepackage[english]{babel} % Language settings
\title{Revolutionizing 6G Wireless Networks with RISs: Enhancements in Sensing and Communications}
\author{Jiguang~He,~\IEEEmembership{Senior Member,~IEEE}, Yuwen~Cao$^*$,~\IEEEmembership{Member,~IEEE}, Charles Vanwynsberghe, ~\IEEEmembership{Member,~IEEE}, Aymen Fakhreddine,~\IEEEmembership{Member,~IEEE}, Yi~Fang,~\IEEEmembership{Senior~Member,~IEEE}, Chongwen Huang,~\IEEEmembership{Member,~IEEE},\\ George C. Alexandropoulos,~\IEEEmembership{Senior Member,~IEEE}, and Mérouane Debbah,~\IEEEmembership{Fellow,~IEEE}
\thanks{The work was supported by the Funding Program for High-Level Overseas-Educated Talents (Grant No. RCXMA26001), the Guangdong Provincial  Project (Grant No. 2024QN11X433), Guangdong Research Team for Communication and Sensing Integrated with Intelligent Computing under Grant 2024KCXTD047, the National Natural Science Foundation of China under Grant 62301143, Shanghai Sci-Tech Co-Research Program under Grant 25HB2702400, the International Collaborative Research Program of Guangdong Science and Technology Department under Grant 2025A0505020081, the Austrian Science Fund (FWF\,--\,Der Wissenschaftsfonds) under grant ESPRIT-54 (Grant DOI: 10.55776/ESP54), and the SNS JU project 6G-DISAC under the European Union's Horizon Europe research and innovation programme under Grant Agreement No. 101139130. (Corresponding author: Yuwen Cao)}
}

\date{\today}

\begin{document}

\maketitle

\begin{abstract}
% Discuss the innovative use of Reconfigurable Intelligent Surfaces (RIS) to significantly improve aspects of modern wireless networks. Highlight the paper's exploration into RIS applications in drone detection, enhanced signal coverage, precise localization, and the leverage of artificial intelligence for real-time RIS control. Emphasize the investigation into maintaining robust RIS connectivity for moving objects.
Reconfigurable intelligent surfaces (RISs) have emerged as one of the most promising technologies for the next-generation wireless networks, offering advantages such as cost efficiency, low power consumption, flexible deployment, and large effective aperture. Extensive studies have demonstrated the capabilities of RISs to enhance communication performance, including coverage extension, sum-rate improvement, interference suppression, security enhancement, etc. As 6G networks continue to evolve, wireless sensing is expected to play a pivotal role alongside communication. Undoubtedly, RISs possess great potential to enable seamless integration of sensing and communication by supporting both passive and active sensing paradigms. This paper explores how RISs can revolutionize wireless networks by enhancing both communication and sensing capabilities. Specifically, we highlight the fundamental principles, potential applications, and future trends of RIS-assisted integrated sensing and communication (ISAC), anticipating that RIS-enabled ISAC will become a key enabler in the upcoming 6G era.  
\end{abstract}


\section{Introduction}
% Introduce the concept of RIS and its transformative potential in wireless networks. Discuss the growing need for enhanced drone detection, signal coverage, and localization accuracy in contemporary networks. Introduce the use of machine learning and deep learning as pivotal to achieving efficient, real-time control of RIS. Present the novel challenge and importance of ensuring RIS connectivity for on-the-move objects.

% \section{Background and State-of-the-Art}
% Provide an overview of RIS technology's evolution, focusing on its emerging roles in wireless communications. Review current methodologies in drone detection, coverage enhancement, and localization via traditional technologies and the limitations therein. Discuss the precedent for using AI in network optimization and control, setting the stage for its application in RIS.

The sixth generation (6G) of wireless networks is envisioned to support immersive communications, massive connectivity, and hyper-reliable low latency wireless links, evolving beyond the three primary usage scenarios of the fifth-generation (5G) networks: enhanced mobile broadband (eMBB), massive machine-type communication (mMTC), and ultra-reliable low-latency communication (URLLC). To achieve this evolution, 6G will face more stringent requirements in key performance indicators (KPIs), including higher spectral efficiency, extremely high peak data rates, centimeter-level positioning accuracy, and sub-millisecond latency. In addition to the evolved versions of the 5G use cases, three new ones have emerged as major driving forces of 6G development: ubiquitous connectivity, native artificial intelligence (AI) communications, and integrated sensing and communication (ISAC)~\cite{Marco2020}.

A transformative enabler for the latter use cases is the technology of reconfigurable intelligent surfaces (RISs)~\cite{RISsurvey2023}, which offers the capability to intelligently manipulate the wireless propagation environment through a large number of cost-efficient, quasi-passive, programmable meta-atoms. RISs can significantly improve system performance supporting tasks such as interference mitigation, signal-to-interference-plus-noise ratio (SINR) enhancement, and network coverage extension, while also paving the way for seamless integration of sensing and communication functions~\cite{10243495}. 

Communications have been the primary service of cellular networks since their inception. Starting from the fourth generation (4G), high accuracy localization has become feasible due to the availability of wider bandwidths and the adoption of the multiple-input multiple-output (MIMO) technology. Since then, continuous advancements have led to increasingly precise localization capabilities, laying the foundation for ISAC~\cite{10872862}. This paradigm allows for the joint optimization of communication and sensing objectives within a unified framework, maximizing the utilization of network resources. The integration of RISs into cellular networks has the potential to further boost ISAC performance by providing unprecedented control over radio wave propagation~\cite{10243495}. For localization and other sensing tasks, RISs, when deployed at carefully selected offline fixed positions, can provide high spatial resolution due to their large aperture and can act as additional positioning reference nodes~\cite{10044963}, thereby complementing conventional base stations (BSs)~\cite{Henk2020}. The integration of uncrewed aerial vehicles (UAVs) into cellular networks has attracted significant attention due to their mobility, flexibility, and ability to provide cost-effective wireless access, particularly in communication-blind or disaster-stricken areas. In this context, RIS-assisted cellular networks can not only significantly enhance the detection, localization, and tracking performance of non-cooperating UAVs, but also actively support UAV self-localization and navigation, thereby enabling reliable communication and sensing for emerging low-altitude economy applications.

Unlike that in \cite{Zhao2026} the authors considered design of RIS-enabled ISAC networks to enable three-dimensional (3D) continuous coverage and ensure seamless network coexistence for UAVs. In this article, we investigate the role of RISs in two cellular network applications involving UAVs: detection and tracking of non-cooperating UAVs as well as UAV self-localization and navigation. We present specific system architectures for both applications as well as experimental results showcasing that RISs can significantly enhance sensing capability. Several challenges related to the RIS deployment, such as the real-time control of their phase profiles, accurate channel estimation, and practical hardware implementation, are identified and promising research directions for both the wireless communications and sensing communities are discussed. 

\section{RIS-Enabled Coverage Enhancement}
One of the most direct and compelling applications of RISs is network coverage enhancement~\cite{Sang2024}. This can be achieved through two primary mechanisms: i) Bypass non-line-of-sight (NLoS) obstructions: RISs can establish alternative reflection paths, allowing communication signals to reach previously blocked areas, such as regions behind buildings or corners. ii) Channel power improvement through intelligent beamforming: By optimizing the RIS beamforming vector, the end-to-end channel gain can be significantly increased, thereby enhancing the overall link quality~\cite{RISsurvey2023}.

\begin{figure}
    \centering
\includegraphics[width=0.9\columnwidth]{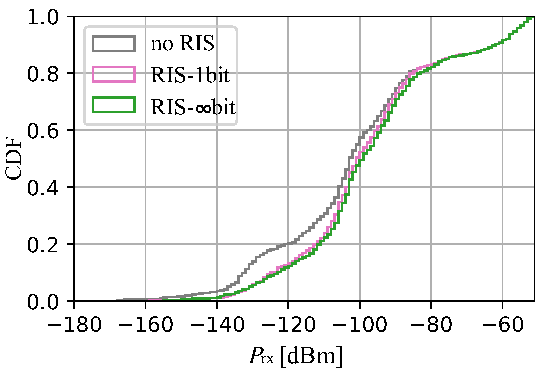}
    \caption{CDF of received signal strength in an RIS-aided cellular network.}
    \label{fig:cdf_rec_sig_str}
    \vspace{-0.7cm}
\end{figure}

To evaluate this potential, a coverage extension study was conducted in a rectangular area in Abu Dhabi near AI Qana with dimension $280 \times 370$ m using Siradel Volcano ray-tracing simulator. The channel is a mixture of Rayleigh and Rician distributions depending on the availability of LoS connectivity. The carrier frequency and bandwidth are 3.5 GHz and 20 MHz. The BS has a single antenna and the RIS has $400$ ($20 \times 20$) elements. The cumulative distribution function (CDF) curves presented in Figure~\ref{fig:cdf_rec_sig_str} are obtained empirically from the received power samples over all spatial grid points in the simulated area, where the wireless channels are generated using the simulator under different RIS configurations. The results demonstrate that a perfect RIS with continuous phase control provides substantial performance gains in the received signal strength, particularly in the transmit power range of $[-140,-100]$ dBm. Even with 1-bit phase quantization, which greatly simplifies hardware implementation, RISs still provide notable performance gains. Moreover, as demonstrated in~\cite{Hongliang2020}, a 2-bit phase quantization is sufficient to achieve performance nearly identical to that of continuous phase control, thereby striking an effective balance between hardware complexity and system performance. By extending coverage, RISs not only enable more users to be served by network operators but also facilitate the provision of additional services in previously inaccessible areas. 

% For instance, localization and sensing applications can be deployed in hard-to-reach locations, unlocking new opportunities for smart city infrastructure and mission-critical services.

\vspace{-0.3cm}
\section{RIS-Enabled UAV Detection and Tracking} \label{sec_drone_detecton_localization}

% {\color{red} Whether we need to include active localization with RIS or not?}
The past decade has witnessed a sharp rise in commercial drone usage, driven by decreasing costs, enhanced functionalities, and greater accessibility to the general public. While these advancements have brought numerous benefits, they also pose significant security and privacy risks, since the potential for UAV misuse or malicious activities has increased. This trend highlights the urgent need for local and national authorities to detect, localize, and track non-cooperating UAVs (i.e., vehicles that do not communicate with the network and may be hostile).

The global deployment of BSs has enabled ubiquitous communication coverage for terrestrial users. However, most BS antennas are down-tilted, primarily designed to serve ground-level devices, leaving the aerial domain insufficiently covered. By strategically deploying RISs in proximity to BSs, the coverage can be effectively extended to the sky, creating a 3D communication and sensing environment. This RIS augmentation allows BSs to support advanced functions, such as 3D localization and tracking of non-cooperating UAVs~\cite{Jiguang2024}. 
%, operating in both passive and active sensing manners
Beyond localization and tracking, RISs also enable other sensing tasks, such as UAV detection and environmental reconstruction. In this section, we first introduce the system architecture of RIS-augmented cellular networks designed for UAV detection, localization, and tracking, and then describe the experimental implementation and validation of RIS-enabled UAV detection using an anechoic chamber setup. %Simulation and experimental results demonstrate that UAVs located in front of the RIS can be accurately detected by analyzing the received signal power spectrum, thereby confirming the feasibility of RIS-assisted aerial sensing.

\vspace{-0.3cm}
\subsection{Cellular System Architecture for Aerial Sensing}
\begin{figure}[!t]  
    \centering
    \includegraphics[width=0.9\columnwidth]{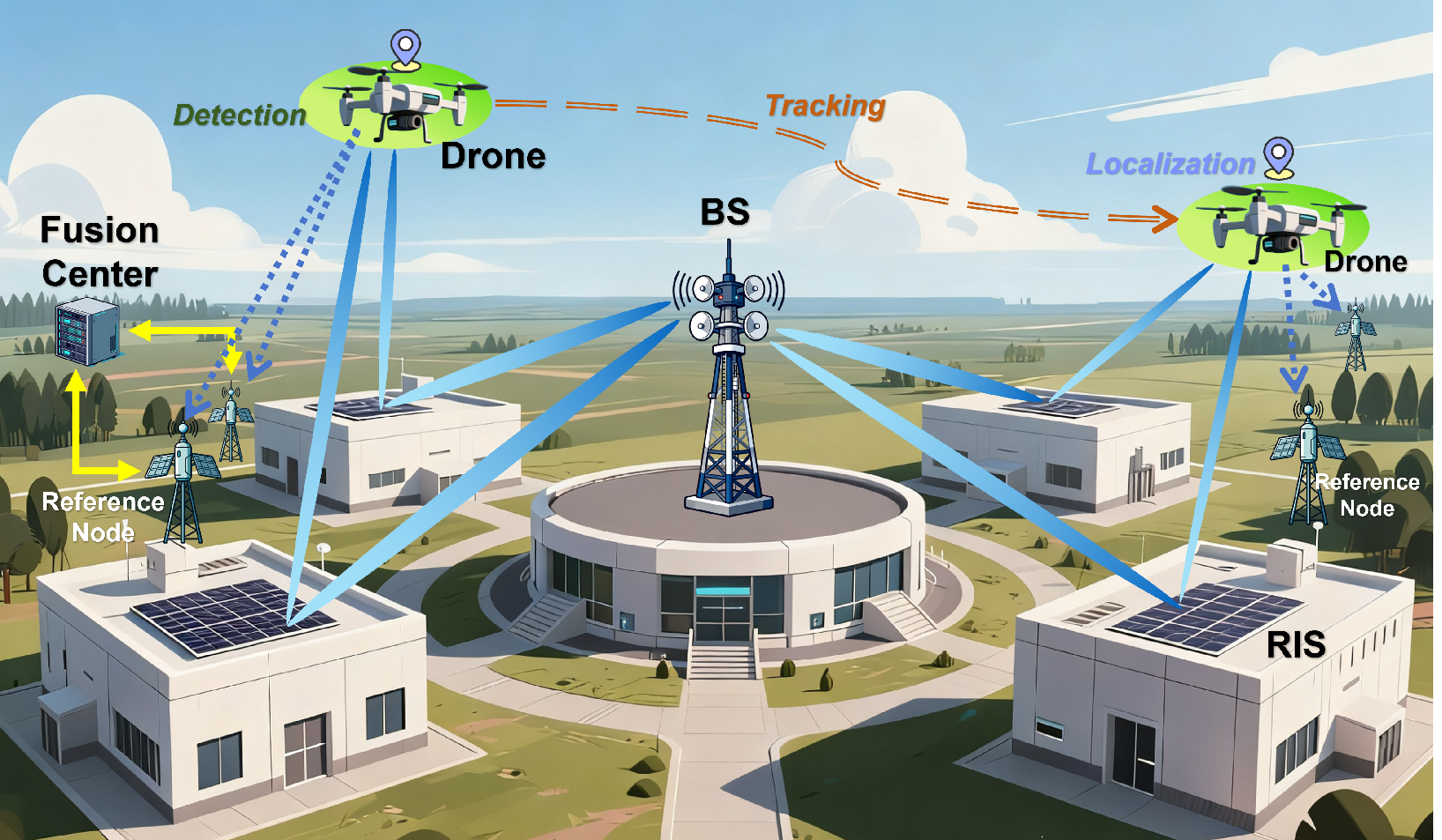}
     \caption{Detection, localization, and tracking of non-cooperating UAVs in RIS-augmented cellular networks.}
       \label{fig:drone_detection}  
     \vspace{-0.7cm}
\end{figure}

Traditional military radar systems are primarily designed to detect large aircraft or UAVs with large radar cross sections (RCSs) operating at high altitudes. However, these legacy systems often fail to provide adequate detection capabilities for small, low altitude UAVs, which have much smaller RCS and more agile flight patterns. Alternative detection methods, such as acoustic arrays and optical cameras, have been explored but are limited by short detection ranges and vulnerability to environmental factors, such as noise, lighting, and weather conditions. To overcome these limitations, an RIS-augmented cellular system architecture for UAV detection, localization, and tracking has been designed, which is illustrated in Figure~\ref{fig:drone_detection}. Unlike the RISs in Section II, which are primarily used for blockage mitigation and coverage extension, the RISs considered here are strategically deployed around cellular BSs to redirect cellular signals toward the sky, thereby creating tunable reflective beams for aerial monitoring. To illuminate the monitored space, a series of RIS reflective beams are designed to uniformly cover the target region, where the beam set can be jointly optimized according to the size and geometry of the monitored area. At the same time, reference nodes are deployed across the coverage area to receive the reflected signals from the UAVs. The placement and orientation of the RISs need to be optimized to maximize the illuminated airspace. The fusion center, functioning as a central processing unit, aggregates the received pilot signals from multiple reference nodes to perform the detection, localization, and tracking of non-cooperating UAVs. By jointly processing pilot signals from multiple reference nodes, the proposed framework can achieve significantly improved sensing performance compared to the single-reference-node case.

Unlike traditional communication scenarios, any direct communication (i.e., information exchange) between the UAV and the cellular network infrastructure is absent. Instead, the system operates similarly to a multi-static radar, where the detection relies purely on passive reflections of the positioning reference signals. When a UAV enters the monitored region, its presence causes an abrupt change in the received power profile at the reference nodes on the monitored frequency bands. In principle, the direct BS signal can be effectively separated from the UAV-reflected components due to the pre-known positions of the BS and reference nodes. In addition, environmental clutter can be mitigated by comparing the pilot signals collected in the presence and absence of such interference. Alternatively, advanced localization algorithms can be employed to jointly estimate the positions of the UAV and surrounding clutter scatterers. By connecting multiple RISs and reference nodes, the system can localize the trespassing UAV by jointly exploiting: (1) angle-of-departure (AoD) information at the RISs which control the direction of the reflected beams; and (2) angle-of-arrival (AoA) information at the reference nodes, which are leveraged to capture the direction of the incoming reflections. Moreover, when the UAV is moving, the RIS beams can dynamically track its movement by continuously adjusting the reflection patterns, thereby ensuring a satisfactory signal-to-noise ratio (SNR) at the receiving nodes for reliable UAV detection and tracking.

% In scenarios where the drone is actively connected to the cellular network, the system enables active 3D localization and real-time tracking. By leveraging pilot signals transmitted from the BS, the drone can extract key channel parameters, such as the AoA information from the RIS reflections and the propagation delay from multiple BSs. Even with limited pilot overhead, this setup allows drones to track their positions in real time.
\vspace{-0.3cm}
\subsection{Simulation Results}
\begin{figure}[t] % Use !h for "here" if possible, but LaTeX might move it  
    \centering
    % Subfigure 1
    \subfloat[Probability of detection with and without using an RIS, where we consider transmission of $90$ pilots for training, set the false alarm probability as $0.001$, and randomly generate the RIS phase profiles.]{
        \begin{minipage}{0.8\columnwidth} % Ensure minipage occupies the full column width
            \centering
            \includegraphics[width=0.8\columnwidth]{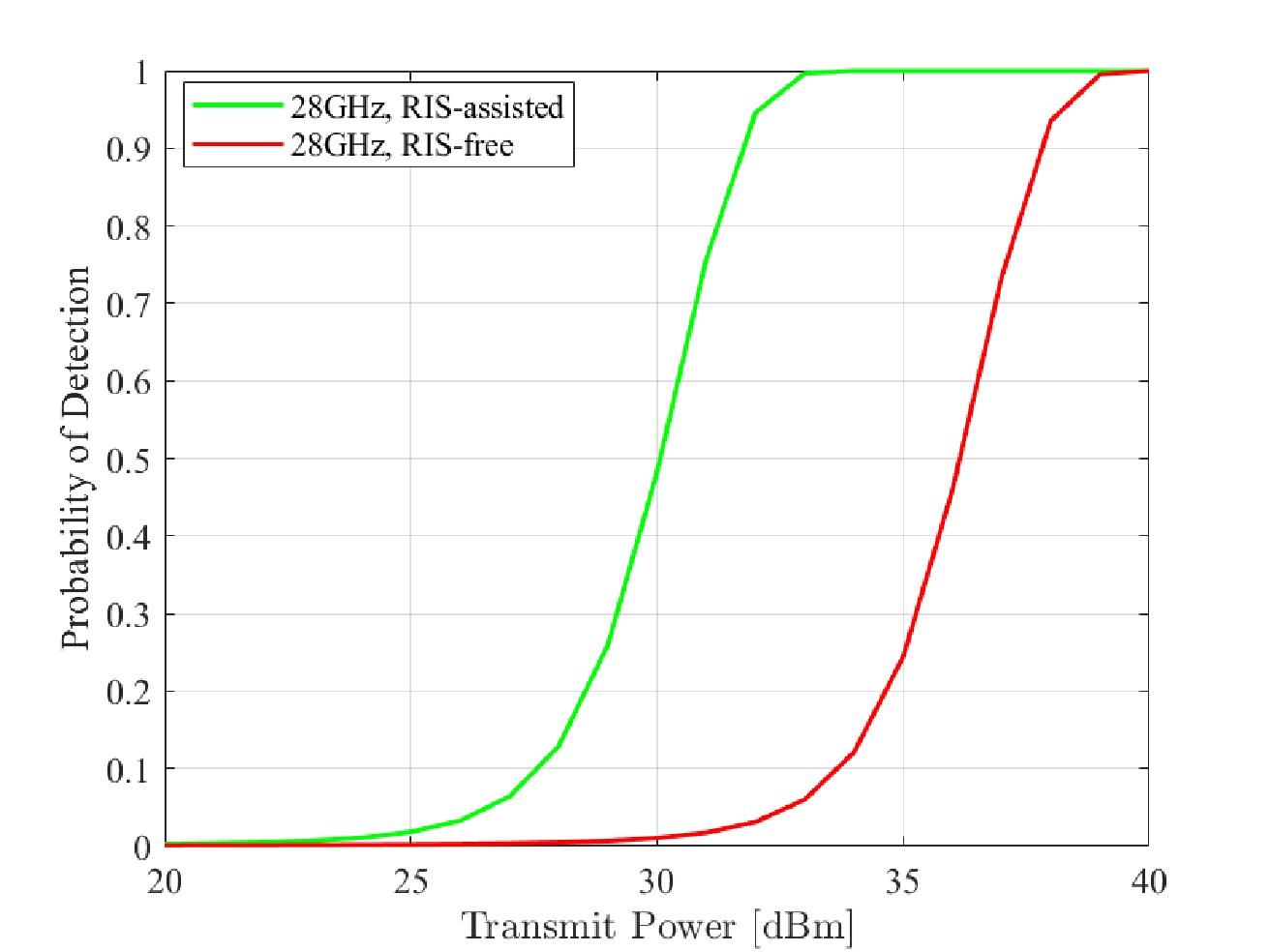} % Adjust width as needed
            \label{fig:subfigA}
        
        \end{minipage}
    }
 
    \par % Force a new line for the next subfigure
    % Subfigure 2
    \subfloat[RMSE performance with and without using an RIS, where we consider transmission of $60$ pilots for training and set RCS as $1$.]{
        \begin{minipage}{0.9\columnwidth} % Ensure minipage occupies the full column width
            \centering
            \includegraphics[width=0.8\columnwidth]{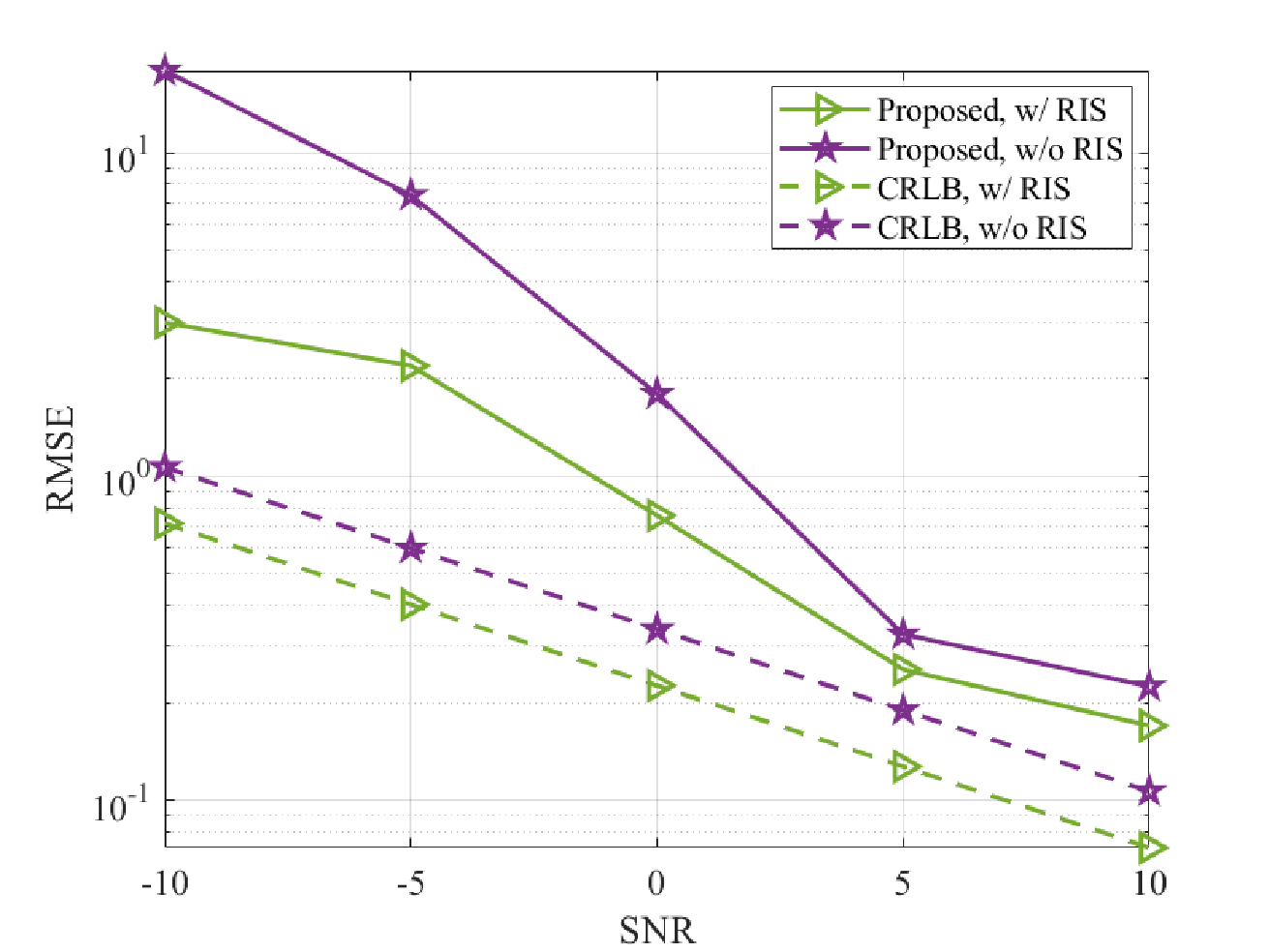} % Adjust width as needed
            \label{fig:subfigB}
           
        \end{minipage}
    }
    \caption{UAV detection and localization with the aid of an RIS operating at $28$~GHz.}
   
    \label{fig:drone_detection_and_localization}
    \vspace{-0.7cm}
\end{figure}

To evaluate the detection and localization performance of non-cooperating drones, we consider a scenario consisting of one BS, one RIS, one non-cooperating drone, and one UE, each placed at fixed three-dimensional positions. The system employs large antenna arrays, operates in the millimeter wave (mmWave) band, and adopts a narrow-band configuration. Free-space path loss is assumed throughout the study, and the transmit power is chosen over a practical operating range. For the UAV detection, the number of BS antennas and RIS elements are 100 and 1600, respectively. The generalized likelihood ratio test (GLRT) is adopted for detection. For the UAV localization, the detailed setups and algorithm development can be referred to \cite{Jiguang2024}. The performance enhancement of UAV detection and localization enabled by RIS integration is illustrated in Figure~\ref{fig:drone_detection_and_localization}. As shown in Figure~\ref{fig:drone_detection_and_localization}(a), a several-dB improvement is observed when the system operates in the mmWave frequency band, demonstrating the RIS's capability to enhance signal strength and coverage. Furthermore, Figure~\ref{fig:drone_detection_and_localization}(b) adopts root mean square error (RMSE) as the target sensing performance metric to compare the passive localization accuracy of our proposed coordinate gradient descent-based algorithm with and without the assistance of an RIS, and shows that the gain becomes even more pronounced in the low SNR regime compared to the high SNR levels. Overall, both theoretical Cram\'er Rao lower bound (CRLB) analyses and practical results confirm that RIS-assisted systems deliver consistent performance improvements across the entire SNR range~\cite{Jiguang2024}. Theoretically, centimeter-level localization accuracy (approximately 10 cm) can be achieved with the aid of RIS at the SNR level of around 10 dB. The localization performance can be further improved through optimized RIS deployment, increased RIS beamforming gain, and enhanced UAV RCS.

\vspace{-0.3cm}
\subsection{Experimental Results}
% To assess the feasibility of RIS-augmented UAV sensing, experiments were conducted at the Directed Energy Research Center (DERC) facilities of the Technology Innovation Institute in Abu Dhabi, UAE. 

Real-world experiments are conducted to assess the feasibility of RIS-augmented aerial sensing. To be specific, all measurements are performed inside an anechoic chamber to minimize unintended reflections, thereby emulating a clean outdoor propagation environment and ensuring controlled, repeatable evaluation conditions. Although the experiments are conducted in an anechoic chamber, the suppressed environmental reflections make the propagation conditions reasonably representative of open-space aerial sensing scenarios, while more comprehensive real-world validations remain part of our future work. The setup includes a $40\times40$ $1$-bit RIS operating at 28 GHz, a signal generator and a horn antenna illuminating the RIS, and a horn antenna connected to a down converter and a universal software radio peripheral (USRP) B210 for the reception. 

\begin{figure*}[t]
    \begin{subfigure}{\textwidth}
    	\centerline{\includegraphics[width=0.75\textwidth]{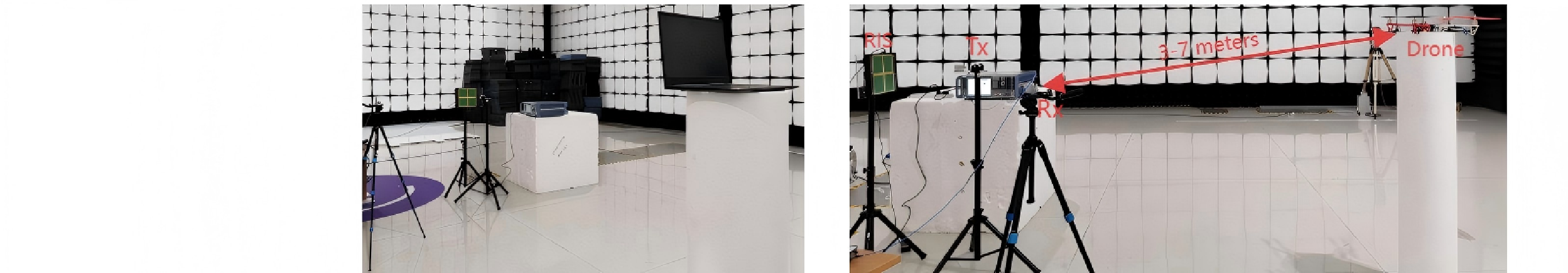}}
    	%\caption{xx}
    	%\label{}
    \end{subfigure}

    \begin{subfigure}{\textwidth}
    	\centerline{\includegraphics[width=0.75\textwidth]{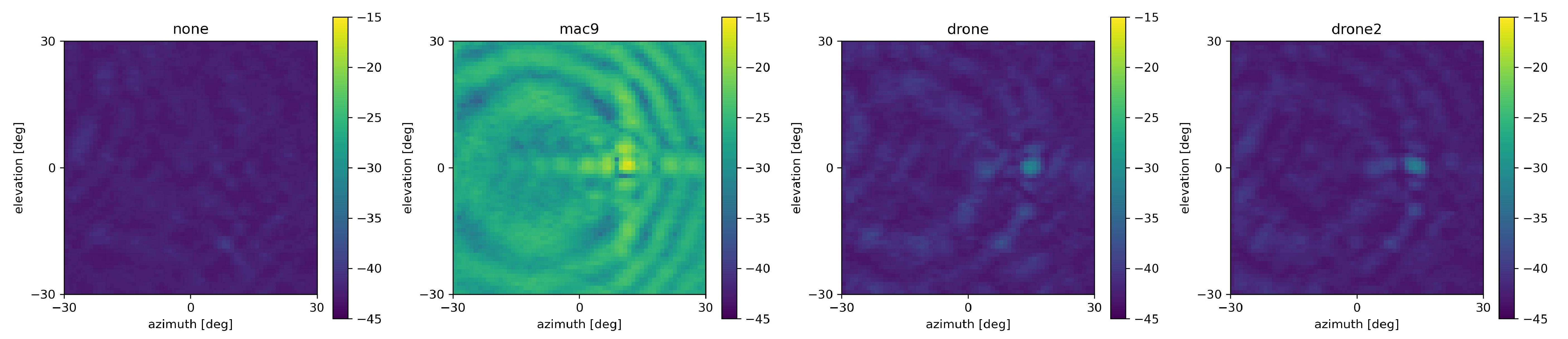}}
    	%\caption{xx}
    	%\label{}
    \end{subfigure}
	\caption{Elevation/azimuth maps of the received power [dB] for different targets at the same location. Targets (from left to right): none, laptop, as well as the UAV with $2$ batteries and when this was covered with copper tape. The maps are plotted on a dynamic range of $30$~dB.
}
		\label{fig:drone_detection_derc}
        \vspace{-0.2cm}
\end{figure*}

\begin{figure*}[t]
    \begin{subfigure}{\textwidth}
    	\centerline{\includegraphics[width=0.75\textwidth]{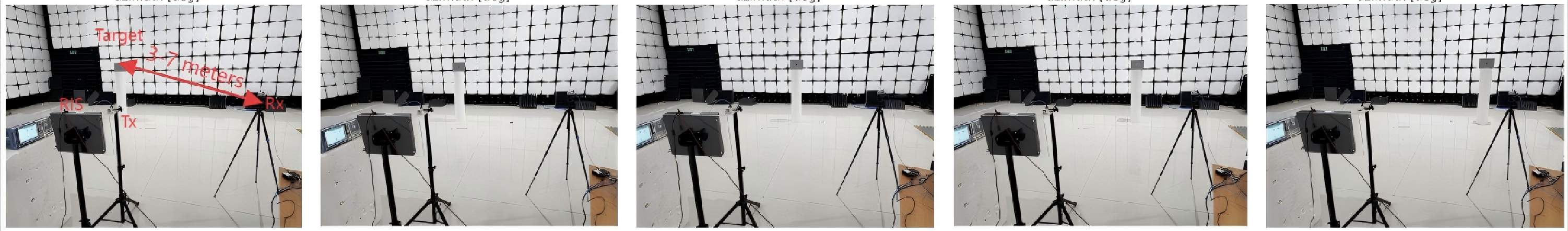}}
    	%\caption{xx}
    	%\label{}
    \end{subfigure}

    \begin{subfigure}{\textwidth}
    	\centerline{\includegraphics[width=0.75\textwidth]{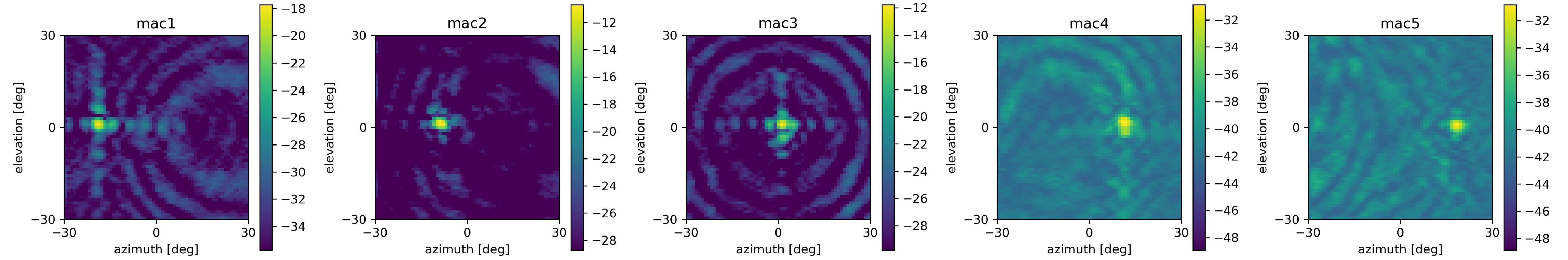}}
    	%\caption{xx}
    	%\label{}
    \end{subfigure}
    \caption{Elevation/azimuth maps of the received power [dB] averaged over $5$ snapshots for laptop target at different locations of the same distance. The maps are plotted on a dynamic range of $18$~dB.}
        \label{fig:localization}
        \vspace{-0.5cm}
\end{figure*}

RCS characterizes the detectability of an object by a radar-like system, where larger RCS values generally indicate stronger reflected electromagnetic energy toward the receiver and are influenced by factors such as the object’s size, material, shape, and orientation. In the experiments, various targets, i.e., a metallic plate (specifically, a laptop), a commercial UAV with batteries, and the same UAV covered with copper tape, are considered at distances of $3–7$ meters from the receiver. Figure~\ref{fig:drone_detection_derc} shows clear differences among their RCSs. Since the BS horn antenna is oriented away from the receiving node, the received signal strength without the RIS is nearly zero, leading to poor detection and localization performance. As observed, the metallic plate produces the strongest reflection, while the drone’s signal is lower but detectable. The copper tape has minimal impact, confirming that the basic UAV structure is sufficient for detection. To test localization of a non-cooperating drone, the metallic plate is moved across five positions. As shown in Figure~\ref{fig:localization}, the BS successfully tracks its movement in the azimuth-elevation plane with the help of an RIS, demonstrating its ability to both detect and locate targets. However, variations in received power across positions reveal that RCS depends heavily on target orientation, limiting the use of power alone for distance estimation.

A key limitation is scanning speed. Scanning a $60^{\circ} \times 60^{\circ}$ area with a $1^{\circ}$ step required $3600$ RIS phase configurations, taking $36$ seconds for one scan and three minutes when averaging $5$ snapshots. Future improvements should focus on hierarchical beam codebooks and faster RIS reconfiguration~\cite{9827873}, enabling practical real-time operation. 

% \begin{figure}[h!]
%     \centering
%     \includegraphics[width=0.6\columnwidth]{figs/drone_detection_experiment.png}
%     \label{fig:drone_detection}
%     \caption{Experimental setup of drone detection and localization}
% \end{figure}

\vspace{-0.2cm}
\section{RIS-Enabled UAV Self-Localization}
\begin{figure}[t]
	\centering
\includegraphics[width=0.99\linewidth]{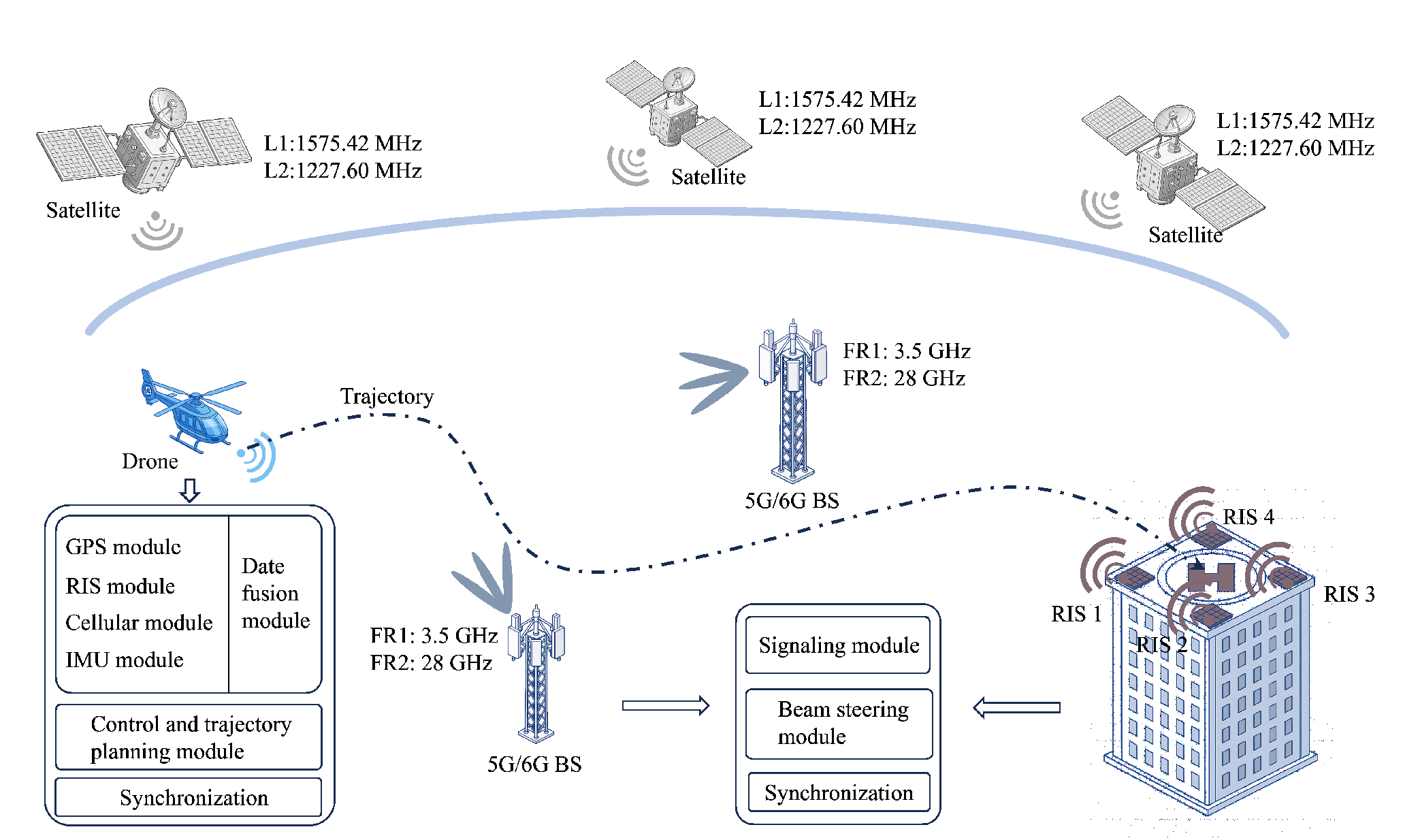}
	\caption{Active drone self-localization and navigation with the aid of RISs and fusion of multiple additional signals, e.g., cellular, RIS, and GPS positioning reference signals.}
		\label{fig:Self_Localization_Drone}
        \vspace{-0.5cm}
\end{figure}

As discussed in Section~\ref{sec_drone_detecton_localization}, the introduction of RISs significantly enhances passive target detection and localization in cellular networks. In such scenarios, sensing and localization tasks are accomplished within a single system without requiring direct communication between the target and the BS. However, when extending to multiple independent localization systems, particularly in outdoor environments where global positioning system (GPS) signals are readily available, the problem setting changes fundamentally. Building on this observation, we shift our focus from non-cooperating UAV to UAV self-localization and navigation, where the drone actively communicates with both satellites and cellular BSs. By leveraging RISs, the system can enhance signal quality and coverage, enabling robust and precise navigation even in challenging environments where conventional GPS or cellular-based solutions alone may be insufficient. Such capabilities are especially relevant for supporting the low altitude economy, including applications such as urban air mobility, aerial logistics, and autonomous drone services.

\vspace{-0.2cm}
\subsection{System Architecture}
Figure~\ref{fig:Self_Localization_Drone} depicts the proposed UAV self-localization and navigation system empowered by RISs and the fusion of multiple independent reference signals. The integration of RIS technology marks a significant advancement in achieving high-precision self-localization and navigation performance. Traditional GPS systems often suffer from signal degradation, multipath effects, and interference, particularly in dense urban environments, leading to reduced accuracy and reliability. To overcome these limitations, the proposed system leverages RIS-assisted signal enhancement combined with state-of-the-art data fusion algorithms, enabling robust and precise localization that goes beyond the capabilities of conventional GPS-only solutions.

% The proposed system consists of a network of RISs—either active or passive—in conjunction with 5G/6G BSs and satellites. The drone, which performs self-localization and navigation, is equipped with multiple independent sensing and communication modules, including a GPS module, a cellular communication module, an RIS interaction module, and an inertial measurement unit (IMU). The outputs from these modules are integrated through a data fusion module, which combines information from heterogeneous sources to provide accurate and reliable position estimation and navigation.

Our unified system at the drone comprises the GPS module, RIS interaction module, cellular communication module, and inertial measurement unit (IMU) module. The outputs from these modules are integrated through a data fusion module, and afterward, the computed results of the proposed low-complexity Kalman filter-based data fusion algorithm are sent into the control and trajectory planning module, to provide accurate and reliable position estimation and navigation. Our unified system is feasible in practice deployment as the mature high-rate lithium-ion batteries have been widely applied on a large scale, and the large-scale computation tasks can be offloaded to the cloud server.

The BSs operate in both FR1 bands (e.g., 3.5 GHz) and FR2 bands (e.g., 28 GHz), supporting high-capacity, low-latency communications, while the satellites provide signals in L1 (e.g., 1575.42 MHz) and L2 (e.g., 1227.60 MHz) bands for traditional GPS-based localization. The RISs play a pivotal role by either transmitting or reflecting pilot signals toward the drone, functioning as active RISs or passive RISs, respectively. By dynamically steering and focusing signals toward the UAV, the RISs enhance signal quality, improve spatial resolution, and provide additional localization measurements, thereby enabling more reliable and accurate UAV navigation in challenging environments.

% By exploiting their capability for super-directional and dynamic beam steering, RISs can focus energy precisely in the desired direction, thereby achieving high spatial resolution and significantly improving localization accuracy.      

% The heart of this enhanced localization methodology lies a sophisticated assembly constituting a reconfigurable intelligent surface (RIS) transceiver and a positioning signal receiver, both integral to the drone’s control apparatus. This system's essence is its dual aptitude to meticulously process signals received from RIS devices and standard positioning systems, subsequently amalgamating these through high-caliber data fusion algorithms. Prominently, the utilization of an evolved variant of the extended Kalman filter stands out, mitigating the congenital inaccuracies induced by urban GPS signal degradation and propelling the localization process towards remarkable precision.

% \paragraph{Operational Dynamics}
% The efficacy of this system is significantly amplified by the dynamic calibration of the RIS transceiver with a spectrum of RIS devices. By harmonizing both passively reflected and actively transmitted RIS signals, the system is adept at constituting a multi-dimensional understanding of the drone's spatial locale. This synergetic signal processing enables leveraging advanced geometric positioning techniques, including trilateration, thereby enhancing the positional accuracy to an unprecedented degree.
\vspace{-0.4cm}
\subsection{Data Fusion for Self-Localization and Navigation}
A key feature of the proposed system is the onboard processor’s ability to synergize signals from diverse sources, namely, RIS-assisted reflections/transmissions, GPS satellites, and cellular BSs, to deliver highly accurate and reliable localization. This integration is achieved through advanced data fusion algorithms that intelligently combine information from these heterogeneous inputs. Among these, an enhanced variant of the extended Kalman filter (EKF) plays a central role. Unlike conventional EKF implementations, the proposed fusion algorithm adaptively adjusts the weights of the RIS-related signals and other modalities for accurate drone localization. For instance, when GPS signals are degraded by multipath effects in dense urban environments, the filter can automatically assign higher confidence to RIS-related or cellular measurements. Conversely, in open areas with strong satellite visibility, GPS inputs may dominate. This continuous adaptation enables the system to seamlessly optimize the drone’s localization in real time, ensuring robust navigation even under challenging conditions such as GPS-denied zones, NLoS scenarios, or rapid drone maneuvers. By fusing RIS signals with traditional positioning systems, the processor effectively compensates for individual weaknesses of each source, ultimately achieving robust centimeter-level localization accuracy.

% \subsection{Implications and Future Prospects}
% The integration of RIS technology with dynamic data fusion algorithms signifies a monumental leap forward in drone self-localization capabilities. By circumventing the limitations inherent to GPS-dependent frameworks in urban environments, the system facilitates drones to undertake a wide array of tasks — ranging from logistical deliveries to intricate emergency response assignments — with an enhanced level of precision and reliability. Beyond its immediate implications for drone navigation, this system leverages the vast potential of RIS in augmenting the functionalities of wireless networks, heralding a new era in telecommunications and robotics.

% \section{Experimental Results and Discussion}
% Present experimental setups, methodologies, and results from the study. Discuss the implications of findings within the broader context of RIS research and potential for future network technologies.

% \section{Challenges, Opportunities, and Future Directions}
\vspace{-0.3cm}
\section{Challenges and Opportunities}
%
% Reflect on the challenges faced during the research, particularly in AI integration and mobile connectivity. Outline opportunities for further research and the potential for RIS technologies to transform future wireless networking landscapes.
In this section, we discuss the potential challenges, particularly in AI integration, mobile connectivity, and multi-mode operation evolution, and then outline opportunities for further research.
\vspace{-0.3cm}
\subsection{AI-Driven RIS Control for Real-Time Adaptation}
%Introduce the integration of machine learning and deep learning techniques for the dynamic control of RIS elements. Present models and algorithms developed for predictive and adaptive RIS behavior, facilitating real-time adjustments to network conditions.
% Building on the earlier introduction of RIS-assisted communication and aerial sensing, achieving real-time responsiveness in such systems hinges critically on intelligent and adaptive RIS control. 

In practical RIS-assisted systems, the RIS placement, orientation, and phase shift configuration should be jointly optimized according to the communication and sensing requirements of different deployment scenarios, while emerging active and sensing-enabled RIS architectures integrated with massive MIMO can alleviate the double path loss effect and further enhance communication and sensing performance.

% The RIS technology has emerged as a wireless coverage enabler, providing high-quality SINR coverage and data rate for mmWave and beyond networks, with low energy consumption, easy deployment, and flexible manipulation of electromagnetic waves. 
% RIS is a planar array composed of periodically arranged subwavelength-sized passive reflecting elements, 
% each with individually controllable reflection coefficients. 
% By simultaneously controlling electromagnetic waves and adjusting the reflection characteristics of these elements, RIS can enhance the received signal quality/strength, 
% mitigating obstacle-induced radio frequency (RF) signal degradation.
% \textcolor{red}{RIS is a planar array of subwavelength-sized passive reflectors with adjustable reflection coefficients, enhancing received signal quality by controlling electromagnetic waves and mitigating RF signal degradation caused by obstacles.}
However, employing RISs will complicate the channel estimation and resource allocation, with its complexity being proportional to the number of RIS elements. Thus, it becomes difficult to facilitate a real-time RIS adaptation to time-variant network conditions, especially in large RIS networks. To tackle the challenges above, one can leverage AI to  dynamically configure a wave-controlled RIS, based on the training data collected by sampling the reflected radiation pattern of the RIS \cite{9864655}. In general, AI-based solutions for dynamic control of RIS elements in the literature can be decomposed into the following categories: 
% i) machine learning (ML)-based, and ii) the integrations of ML and heuristic solutions, which will be respectively introduced as follows:
% \begin{itemize}
    % \item \textbf{Model-based algorithms}: Model-based algorithms combine theoretical frameworks from traditional optimization algorithms with deep learning models. To accelerate the convergence, such kind of algorithms unfold the iterative optimization algorithm. However, they will incur huge training and feedback overhead of acquiring channel state information (CSI) with passive RIS, and are invalid to the scenario where system CSI is unknown. In addition, to improve feasibility for real-time implementation, model-based methods should better take into account factors such as frequency dependence or device nonlinearities. 
    %
    % \item \textbf{Data-based algorithms}: Data-based methods usually rely on large amounts of statistical data for neural model training. However, the accuracy performance of data-based methods is sensitive to the quality of the ground truth data. Besides, there is also difficulty in architecting the neural networks for system modeling while avoiding overfitting, with some solutions including adding dropout layers (randomly ignoring neuron contributions), using larger datasets, and decreasing the number of hidden layers.
    %
i) \textbf{Machine learning (ML)-based algorithms}: Advanced ML methods, e.g., deep neural network, transformer, computer vision methods, have been implemented into RIS-assisted networks to distinguish line-of-sight (LoS) link and NLoS link to precisely predict channel information, as well as to aid RIS for dynamic beam tracking and controlling the reconfiguration of the RIS's phase shifts to account for rapid changes in dynamic channels. In particular, one representative learning-based algorithm was proposed in \cite{10930892}, which presented a novel multi-branch attention convolutional neural network architecture designed for realistic RIS models with discrete state elements. This algorithm can jointly control the configuration of the RIS's discrete phase responses and the transmit precoder in real-time manner. In addition, “learning-to-control” paradigm shows great benefits in realizing the near-optimal computation and real-time adaptation to time-varying network conditions when dealing with the resource allocation and RIS phase shifting control tasks in RIS-assisted networks. ii) \textbf{Hybrid ML-based and heuristic algorithms}: 
Such kind of hybrid algorithms can take the advantages of both ML-based and heuristic solutions while overcoming their drawbacks. Specifically, ML-based algorithms have been proven effective in capturing the relationships between RIS configurations and achievable rates at arbitrary UE’s locations. Meanwhile, an optimal set of discrete RIS phase shifts was recommended to optimize the propagation environment. This is done by an exhaustive search on an existing codebook of RIS phase shifts during inference, which adds additional delays.   
% \end{itemize}
\vspace{-0.3cm}
\subsection{On-the-Move RIS Connectivity}
As shown in the drone self-localization and navigation scenarios above, RIS-enabled systems must reliably support objects in motion, making mobility a fundamental challenge for future RIS-assisted ISAC applications. Mobility requires real-time RIS reconfiguration, which can cause latency and signaling overhead if it is not efficiently managed. Ensuring continuous connectivity with mobile users is challenging due to rapid position changes, frequent blockages, and dynamic channel conditions, leading to sudden performance drops. To address these issues, ML-based prediction is used to anticipate user movement and proactively adjust RIS beams. Hierarchical codebook design and beam management help reduce reconfiguration time~\cite{9827873}, while cooperative RIS clusters provide overlapping coverage for seamless handovers. For aerial sensing, mobility adds complexity through Doppler shifts and fast-changing reflections, affecting accurate target tracking. ISAC improves beam control and sensing reliability, ensuring uninterrupted service and enhanced network performance even in highly mobile environments.
\vspace{-0.3cm}
\subsection{Multi-Functional RIS}
The drone self-localization and navigation introduced earlier motivates the need for more flexible RIS architectures, among which multi-functional RIS stands out as a key enabler for next-generation ISAC systems. Multi-functional RIS extends the concept of conventional RIS by supporting multiple operating modes simultaneously, such as transmission and reflection.  Among them, one of the most widely studied is the simultaneously transmitting and reflecting RIS (STAR-RIS)~\cite{9570143}. This dual-mode capability allows a single outdoor BS to provide seamless coverage for both indoor and outdoor users when the STAR-RIS is strategically deployed, for instance, on building windows. Such a setup can greatly improve coverage in areas where direct BS signals are blocked, enhancing overall network efficiency.

However, these benefits introduce significant challenges. The most critical issue is the design of coupled control matrices for transmission and reflection modes, as the two processes are inherently interdependent. This coupling complicates beamforming design, channel estimation, and real-time reconfiguration, especially in dynamic environments with mobile users. In the context of sensing, STAR-RIS enables simultaneous indoor and outdoor localization, opening new possibilities for ISAC. Yet, the challenge lies in optimizing the two control matrices jointly while managing limited network resources such as spectrum allocation, transmit power control, and power-splitting coefficients. Future research should focus on scalable resource allocation algorithms, low-complexity hardware architectures, and ISAC-centric designs that fully exploit multi-function capabilities.

% \subsection{Holographic MIMO}

% \subsection{\textcolor{red}{Future Directions}}
\vspace{-0.3cm}
\section{Future Research Directions}
To fully leverage RIS technology in 6G, it is essential to explore its applications in over-the-air (OTA) computation, generative AI, and distributed intelligent sensing and communications (DISAC). This section introduces key research topics that deserve further exploration.
\vspace{-0.3cm}
\subsection{Deep Over-the-Air Computation via RIS}
Earlier discussions on RIS-assisted coverage enhancement and aerial sensing demonstrate the value of RIS in shaping wireless propagation, and a natural extension is to exploit this capability for computation-oriented tasks in 6G networks~\cite{RISsurvey2023}. With the emergence of artificial intelligence of things (AIoT), the exponential growth of data volume, complexity, and latency requirements introduces major challenges. To overcome the aforementioned challenges, OTA computation as a promising technique integrating the computational process into the communication framework has emerged to facilitate fast data aggregation and processing \cite{10077112}. OTA computation together with RIS can enlarge the wireless coverage, mitigate interference, and enhance the energy- and spectrum-efficiency for MIMO networks. However, how to enforce real-time data aggregation and processing while mitigating communication and computational latency in RIS-assisted multi-cell MIMO networks is unexplored yet. Designing advanced deep OTA computation architecture that is applicable to the RIS-assisted multi-cell MIMO networks is imperative as well.
\vspace{-0.3cm}
\subsection{
Generative AI for Low-Overhead CSI Feedback via RIS}
Since large-scale RISs are implemented without baseband signal processing capabilities, designing their reflection/refraction coefficients requires limited channel state information (CSI) feedback, particularly in frequency division duplex (FDD) massive MIMO-OFDM systems. Existing deep autoencoder (DAE) approaches reduce the feedback overhead and complexity by designing advanced encoder- and decoder-networks. However, these methods rely heavily on large-scale data for model training, thus rendering excessive bandwidth usage and data privacy issues. Besides, the existing CSI feedback model may become outdated when users move to new environments. To address these issues, one can leverage the generative AI to facilitate the CSI feedback training with low-overhead while preserving users’ privacy in RIS-assisted MIMO-OFDM systems.
\vspace{-0.3cm}
\subsection{RIS Empowered ISAC/DISAC}
%
% Joint beamforming and RIS reflection/refraction coefficient designs while guaranteeing superior sensing performance is challenging, particularly in large RIS networks or dynamic environments. To overcome this issue, it is imperative to design model-driven deep learning framework to jointly optimize the transmit beamforming, waveform and RIS reflection/refraction coefficients in ISAC networks with low-complexity.

Interference management is a critical issue in RIS-enabled ISAC networks due to the coexistence of communication and sensing functionalities, where advanced beamforming, precoding, and RIS phase shift optimization techniques can be employed to mitigate multi-user and communication-sensing interference~\cite{10243495}. DISAC extends the capabilities of ISAC, offering an enhanced vision that upgrades the functionality of existing ISAC systems \cite{10872862}. More concretely, DISAC addresses the limitations of current ISAC systems by: (i) enabling widely distributed deployments to track both UEs and sensing targets; 
% (ii) adopting key performance indicators (KPIs) and key value indicators (KVIs) as dual performance metrics; 
and (ii) fusing heterogeneous and distributed sensors to achieve high-resolution UE and target tracking with reduced energy consumption. However, key bottlenecks remain in realizing the DISAC vision, particularly the need for further research on seamlessly integrating semantic communications into the DISAC framework to enhance sensing and communication performance and robustness.  In addition, the integration of multi-functional RISs introduces further complexity, particularly when RISs lack local sensing capabilities.
%

% \subsection{Covert Communications Aided by RIS}
% %
% Given RIS's ability to manipulate propagation for both communication and sensing, as shown from Section II to Section IV, extending these capabilities to covert communication is a promising direction. Existing RIS-assisted covert transmission designs for MIMO ISAC systems may become infeasible in practical settings. This is because most of them either take into account a simple communication model or assume deterministic target location parameters \cite{11062501}. Research focusing on more complex communication model and/or the sensing model with randomly located or location unknown targets is unexplored yet. Besides, it is worthwhile exploring how to design covert beamforming and active/passive RIS reflection coefficients to maximize the secrecy communication rate subject to the communication quality-of-service (QoS), target localization, and power constraints.
%
% \subsection{RIS Empowered DISAC}
%

\vspace{-0.3cm}
\section{Conclusion}
In this paper, we have investigated performance enhancements in both sensing and communication enabled by RISs. Specifically, we have analyzed received signal strength, and device-free drone detection and localization, with validation through real-world experimental measurements. Building upon these results, we have further extended our study to drone self-localization and navigation, leveraging RISs in combination with other sensing modalities. Finally, we have discussed the key challenges faced by RIS-augmented cellular networks and highlight opportunities and future research directions to fully unlock the potential of RISs in the next-generation 6G systems.

% \section*{Acknowledgments}

%\printbibliography

\vspace{-0.3cm}
\bibliographystyle{IEEEtran}
\bibliography{IEEEabrv,Ref}

\begin{thebibliography}{10}
\providecommand{\url}[1]{#1}
\csname url@samestyle\endcsname
\providecommand{\newblock}{\relax}
\providecommand{\bibinfo}[2]{#2}
\providecommand{\BIBentrySTDinterwordspacing}{\spaceskip=0pt\relax}
\providecommand{\BIBentryALTinterwordstretchfactor}{4}
\providecommand{\BIBentryALTinterwordspacing}{\spaceskip=\fontdimen2\font plus
\BIBentryALTinterwordstretchfactor\fontdimen3\font minus \fontdimen4\font\relax}
\providecommand{\BIBforeignlanguage}[2]{{%
\expandafter\ifx\csname l@#1\endcsname\relax
\typeout{** WARNING: IEEEtran.bst: No hyphenation pattern has been}%
\typeout{** loaded for the language `#1'. Using the pattern for}%
\typeout{** the default language instead.}%
\else
\language=\csname l@#1\endcsname
\fi
#2}}
\providecommand{\BIBdecl}{\relax}
\BIBdecl

\bibitem{Marco2020}
M.~Giordani, M.~Polese, M.~Mezzavilla, S.~Rangan, and M.~Zorzi, ``Toward {6G} networks: Use cases and technologies,'' \emph{{IEEE} Commun. Mag.}, vol.~58, no.~3, pp. 55--61, 2020.

\bibitem{RISsurvey2023}
E.~Basar, G.~C. Alexandropoulos, Y.~Liu, Q.~Wu, S.~Jin, C.~Yuen, O.~Dobre, and R.~Schober, ``Reconfigurable intelligent surfaces for {6G}: {E}merging hardware architectures, applications, and open challenges,'' \emph{IEEE Veh. Technol. Mag.}, vol.~19, no.~3, pp. 27--47, 2024.

\bibitem{10243495}
S.~P. Chepuri, N.~Shlezinger, F.~Liu, G.~C. Alexandropoulos, S.~Buzzi, and Y.~C. Eldar, ``Integrated sensing and communications with reconfigurable intelligent surfaces: From signal modeling to processing,'' \emph{IEEE Signal Process. Mag.}, vol.~40, no.~6, pp. 41--62, 2023.

\bibitem{10872862}
E.~Calvanese~Strinati, G.~C. Alexandropoulos, N.~Amani, M.~Crozzoli, G.~Madhusudan, S.~Mekki, F.~Rivet, V.~Sciancalepore, P.~Sehier, M.~Stark, and H.~Wymeersch, ``Toward distributed and intelligent integrated sensing and communications for 6{G} networks,'' \emph{IEEE Wireless Commun.}, vol.~32, no.~1, pp. 60--67, 2025.

\bibitem{10044963}
K.~Keykhosravi, B.~Denis, G.~C. Alexandropoulos, Z.~S. He, A.~Albanese, V.~Sciancalepore, and H.~Wymeersch, ``Leveraging {RIS}-enabled smart signal propagation for solving infeasible localization problems: Scenarios, key research directions, and open challenges,'' \emph{IEEE Veh. Technol. Mag.}, vol.~18, no.~2, pp. 20--28, 2023.

\bibitem{Henk2020}
H.~Wymeersch, J.~He, B.~Denis, A.~Clemente, and M.~Juntti, ``Radio localization and mapping with reconfigurable intelligent surfaces: Challenges, opportunities, and research directions,'' \emph{{IEEE} Veh. Technol. Mag.}, vol.~15, no.~4, pp. 52--61, 2020.

\bibitem{Zhao2026}
Y.~Zhao, M.~Jian, and Y.~Yuan, ``{RIS-Enabled UAV} communications and sensing: Opportunities, challenges, and key technologies,'' \emph{IEEE Trans. Cogn. Commun. Netw.}, vol.~12, pp. 5451--5472, 2026.

\bibitem{Sang2024}
J.~Sang, Y.~Yuan, W.~Tang, Y.~Li, X.~Li, S.~Jin, Q.~Cheng, and T.~J. Cui, ``Coverage enhancement by deploying {RIS} in {5G} commercial mobile networks: Field trials,'' \emph{IEEE Wirel. Commun.}, vol.~31, no.~1, pp. 172--180, 2024.

\bibitem{Hongliang2020}
H.~Zhang, B.~Di, L.~Song, and Z.~Han, ``Reconfigurable intelligent surfaces assisted communications with limited phase shifts: How many phase shifts are enough?'' \emph{{IEEE} Trans. Veh. Technol.}, vol.~69, no.~4, pp. 4498--4502, 2020.

\bibitem{Jiguang2024}
J.~He, C.~Vanwynsberghe, H.~Chen, C.~Huang, and A.~Fakhreddine, ``Device-free {3D} drone localization in {RIS}-assisted mm{W}ave {MIMO} networks,'' in \emph{Proc. IEEE Global Commun. Conf.}, Cape Town, South Africa, 2024, pp. 4436--4441.

\bibitem{9827873}
G.~C. Alexandropoulos, V.~Jamali, R.~Schober, and H.~V. Poor, ``Near-field hierarchical beam management for {RIS}-enabled millimeter wave multi-antenna systems,'' in \emph{Proc. IEEE Sensor Array Multichannel Signal Process. Workshop}, Trondheim, Norway, 2022, pp. 460--464.

\bibitem{9864655}
G.~C. Alexandropoulos, K.~Stylianopoulos, C.~Huang, C.~Yuen, M.~Bennis, and M.~Debbah, ``Pervasive machine learning for smart radio environments enabled by reconfigurable intelligent surfaces,'' \emph{Proc. IEEE}, vol. 110, no.~9, pp. 1494--1525, 2022.

\bibitem{10930892}
G.~Stamatelis, K.~Stylianopoulos, and G.~C. Alexandropoulos, ``Evolving multi-branch attention convolutional neural networks for online {RIS} configuration,'' \emph{IEEE Trans. Cogn. Commun. Netw.}, early access, 2025.

\bibitem{9570143}
X.~Mu, Y.~Liu, L.~Guo, J.~Lin, and R.~Schober, ``Simultaneously transmitting and reflecting ({STAR}) {RIS} aided wireless communications,'' \emph{{IEEE} Trans. Wireless Commun.}, vol.~21, no.~5, pp. 3083--3098, 2022.

\bibitem{10077112}
X.~Li, Y.~Gong, K.~Huang, and Z.~Niu, ``Over-the-air integrated sensing, communication, and computation in {IoT} networks,'' \emph{IEEE Wireless Commun.}, vol.~30, no.~1, pp. 32--38, 2023.

\end{thebibliography}
\vspace {-1cm}
\begin{IEEEbiographynophoto}{Jiguang~He}
(jiguang.he@gbu.edu.cn) is an associate professor at the Great Bay University, Dongguan, China. His research interests span millimeter wave MIMO communications, reconfigurable intelligent surfaces for ISAC, and advanced signal processing techniques.
\end{IEEEbiographynophoto}

\vspace {-1cm}
\begin{IEEEbiographynophoto}{Yuwen~Cao}
(ywcao@dhu.edu.cn) is currently a Lecturer with the School of Information and Intelligent Science, Donghua University, Shanghai, China. His current research focuses on reconfigurable intelligent surfaces, MIMO communications, ISAC, and machine learning.
\end{IEEEbiographynophoto}

\vspace {-1cm}
\begin{IEEEbiographynophoto}{Charles~Vanwynsberghe}
(charles.vanwynsberghe@outlook.fr) is a Senior Researcher at the Technology Innovation Institute, Abu Dhabi. His research focuses on technological enablers for 6G wireless communications and real-time monitoring through RF and underwater acoustic sensing.
\end{IEEEbiographynophoto}

\vspace {-1cm}
\begin{IEEEbiographynophoto}{Aymen Fakhreddine}
(aymen.fakhreddine@aau.at) is a principal investigator at the University of Klagenfurt (Austria) as a recipient of the ESPRIT grant from FWF - Der Wissenschaftsfonds (Austrian Science Fund). His expertise is in wireless communications, networking, and localization.
% He previously worked as senior researcher at the Technology Innovation Institute (UAE), researcher at IMDEA Networks Institute (Spain), visiting researcher at Singapore University of Technology and Design, and research intern at Nokia Bell Labs (France).
\end{IEEEbiographynophoto}

\vspace {-1cm}
\begin{IEEEbiographynophoto}{Yi~Fang}
(fangyi@gdut.edu.cn) is currently a full professor and the vice dean with the School of Information Engineering, Guangdong University of Technology, China. His current research interests include information and coding theory, digital modulation, and cooperative communication. 
\end{IEEEbiographynophoto}

\vspace {-1cm}
\begin{IEEEbiographynophoto}{Chongwen Huang}
(chongwenhuang@zju.edu.cn) is currently a tenure-track Young Professor at Zhejiang University. His main research interests are focused on holographic MIMO surface/reconfigurable intelligent surface, B5G/6G wireless communications, mmWave/THz communications, and deep learning technologies for wireless communications.
\end{IEEEbiographynophoto}

\vspace {-1cm}
 \begin{IEEEbiographynophoto}{George C. Alexandropoulos}
(alexandg@di.uoa.gr) is currently an Associate Professor with the Department of Informatics and Telecommunications, School of Sciences, National and Kapodistrian University of Athens (NKUA), Greece. His research interests span the general areas of algorithmic design and performance analysis for wireless networks with emphasis on multi-antenna transceiver hardware architectures, full duplex radios, active and passive Reconfigurable Intelligent Surfaces (RISs), integrated communications and sensing, millimeter wave and THz communications, as well as distributed machine learning algorithms. 
\end{IEEEbiographynophoto}

\vspace {-1cm}
\begin{IEEEbiographynophoto}{Mérouane Debbah}
(merouane.debbah@ku.ac.ae) is currently a Professor at Khalifa University of Science and Technology in Abu Dhabi and founding Director of the KU 6G Research Center. He is known for his work on Large Language Models, distributed AI systems for networks, and semantic communications.
\end{IEEEbiographynophoto}

\end{document}